\documentclass[aps,prl,superscriptaddress,preprintnumbers,twocolumn,groupedaddress,showpacs,nofootinbib]{revtex4}

\usepackage{axodraw4j}
\usepackage{graphicx}

\def\mathswitchr#1{\relax\ifmmode{\mathrm{#1}}\else$\mathrm{#1}$\fi}

\newcommand{\Pb}{\mathswitchr b}
\newcommand{\Pt}{\mathswitchr t}
\newcommand{\Pu}{\mathswitchr u}
\newcommand{\Pd}{\mathswitchr d}

\newcommand{\Pg}{\mathswitchr g}

\newcommand{\PW}{\mathswitchr W}

\def\mathswitch#1{\relax\ifmmode#1\else$#1$\fi}

\newcommand{\TeV}{\unskip\,\mathrm{TeV}}
\newcommand{\GeV}{\unskip\,\mathrm{GeV}}

\def\refeq#1{\mbox{(\ref{#1})}}

\def\reffi#1{\mbox{Fig.~\ref{#1}}}

\newcommand{\rT}{\mathrm T}
\newcommand{\ri}{\mathrm i}
\newcommand{\rd}{\mathrm d}

\def\eqintext#1{#1}

\newcommand{\cs}{\mathcal{S}}
\newcommand{\calA}{\mathcal{A}}
\newcommand{\calC}{\mathcal{C}}
\newcommand{\calI}{\mathcal{I}}
\newcommand{\calK}{\mathcal{K}}
\newcommand{\calM}{\mathcal{M}}
\newcommand{\calN}{\mathcal{N}}
\newcommand{\calO}{\mathcal{O}}
\newcommand{\calW}{\mathcal{W}}

\newcommand{\col}{\mathrm{col}}
\newcommand{\hel}{\mathrm{hel}}
\newcommand{\re}{\mathrm{Re}}

\newcommand{\X}{X}
\newcommand{\Y}{Y}
\newcommand{\Z}{Z}

\newcommand{\vrad}{21.28}
\newcommand{\radb}{40.0}
\newcommand{\dx}{20}
\newcommand{\dy}{20}
\newcommand{\ysh}{10}

\newcommand{\treereclhs}{
\begin{picture}(40,60)(-13,-30)
\Line(-15,0)(0,0)
\Vertex(-15,0){1.5}
\GCirc(10,0){13}{1}
\put(8,-3){\footnotesize{$i$}}
\end{picture}}

\newcommand{\treerecrhs}{
\begin{picture}(60,80)(-20,-40)
\Line(-20,0)(0,0)
\Vertex(-20,0){1.5}
\Vertex(0,0){1.5}
\Line(0,0)(\dx,\dy)
\GCirc(\dx,\dy){13}{1}
\put(17,18){\footnotesize{$k$}}
\Line(0,0)(\dx,-\dy)
\GCirc(\dx,-\dy){13}{1}
\put(16,-24){\footnotesize{$j$}}
\end{picture}}

\newcommand{\npointloop}{
\begin{picture}(90,104)(-40,-52)
\put(-10,20){\tiny{${n-1}$}}
\put(-25,-1){\tiny{${0}$}}
\put(-3,-24){\tiny{${1}$}}
\ArrowArc(0,0)(30,35,325)
\DashCArc(0,0)(30,325,35){3}
\Line(\vrad,\vrad)(\radb,\radb)
\GCirc(\radb,\radb){13}{1}
\Vertex(\vrad,\vrad){1.5}
\put(30,\radb){\tiny{$i_{n-1}$}}
\Line(-\vrad,\vrad)(-\radb,\radb)
\GCirc(-\radb,\radb){13}{1}
\Vertex(-\vrad,\vrad){1.5}
\put(-45,\radb){\tiny{$i_{n}$}}
\Line(\vrad,-\vrad)(\radb,-\radb)
\GCirc(\radb,-\radb){13}{1}
\Vertex(\vrad,-\vrad){1.5}
\put(36,-\radb){\tiny{$i_{2}$}}
\Line(-\vrad,-\vrad)(-\radb,-\radb)
\GCirc(-\radb,-\radb){13}{1}
\Vertex(-\vrad,-\vrad){1.5}
\put(-44,-\radb){\tiny{$i_1$}}
\end{picture}}

\newcommand{\loopreclhs}{
\begin{picture}(70,50)(-30,-25)
\Line(-30,\ysh)(10,\ysh)
\Line(-30,-\ysh)(10,-\ysh)
\GCirc(20,0){20}{1}
\put(14,-3){{$\calI_{n}$}}
\ArrowLine(-29,\ysh)(-30,\ysh)
\ArrowLine(-30,-\ysh)(-29,-\ysh)
\end{picture}}

\newcommand{\looprecrhs}{
\begin{picture}(96,50)(-50,-25)
\Line(-50,\ysh)(10,\ysh)
\Line(-50,-\ysh)(10,-\ysh)
\GCirc(20,0){20}{1}
\Vertex(-25,\ysh){1.5}
\Line(-25,\ysh)(-25,40)
\GCirc(-25,35){13}{1}
\put(-30,35){\tiny{$i_{n}$}}
\put(8,-3){{$\calI_{n-1}$}}
\ArrowLine(-49,\ysh)(-50,\ysh)
\ArrowLine(-50,-\ysh)(-49,-\ysh)
\end{picture}}

\begin{document}

\preprint{ZU-TH 23/11, LPN11-66}
\title{\boldmath{
Scattering Amplitudes with Open Loops
}}

\author{F.~Cascioli}
\affiliation{Institut f\"ur Theoretische Physik, Universit\"at Z\"urich, 8057 Z\"urich, Switzerland}

\author{P.~Maierh\"ofer}
\affiliation{Institut f\"ur Theoretische Physik, Universit\"at Z\"urich, 8057 Z\"urich, Switzerland}

\author{S.~Pozzorini}
\affiliation{Institut f\"ur Theoretische Physik, Universit\"at Z\"urich, 8057 Z\"urich, Switzerland}

\date{\today}

\begin{abstract}

We introduce a new technique to generate scattering amplitudes at one loop.
Traditional tree algorithms, which handle diagrams with fixed momenta,
are promoted to generators of loop-momentum polynomials
that we call open loops. Combining open loops with
tensor-integral and OPP reduction results in 
a fully flexible, very fast, and numerically stable one-loop generator.
As demonstrated with non-trivial applications, the open-loop approach
will permit to obtain precise predictions for a
very wide range of collider processes.
\end{abstract}

\maketitle


Theoretical simulations of scattering processes play a key role for the
interpretation of data collected at the Large Hadron Collider (LHC). 
Whenever theory predictions are used to link data to model parameters, or to
separate signals from backgrounds, perturbative calculations beyond leading
order (LO) are indispensable, in order to reduce theoretical errors and to
quantify them in a reliable way.
The vast physics programme of the LHC requires next-to-leading-order (NLO)
predictions for a large variety of processes and theoretical models.  In
this context, the fairly large particle multiplicities resulting from the
high collider energy can lead to one-loop amplitudes of unmanageable
complexity.
Handling $2\to 4$ processes with traditional one-loop techniques 
yields severe numerical instabilities and gigantic algebraic expressions,
and can require huge CPU and human power.

The importance of these challenges, marked by the creation of the 2005 Les
Houches priority list~\cite{Binoth:2010ra}, triggered a series
of recent  theoretical developments that led  to the completion of various
multi-particle NLO calculations~\cite{Ellis:2011cr}. By using
tensor-integral reduction and Feynman diagrams, it became possible to handle
multi-particle processes with high efficiency and numerical
stability~\cite{Bredenstein:2009aj, Denner:2010jp}.  Alternatively, new
reductions of on-shell type were
introduced~\cite{Ossola:2006us,Giele:2008ve,Berger:2008sj} that avoid tensor integrals and
reduce all process-dependent aspects of one-loop calculations to a
LO problem.  In this framework, the Ossola-Papadopoulos-Pittau (OPP)
technique~\cite{Ossola:2006us} led to the development of highly automatic
NLO generators~\cite{vanHameren:2009dr,Hirschi:2011pa,Cullen:2011ac}.

One of the features emerging from first LHC applications is
a trade-off between CPU efficiency and automation. While the tensor-reduction approach
leads to the fastest numerical codes~\cite{Bredenstein:2009aj, Denner:2010jp},
at present its large-scale applicability
is limited by the occurrence of very large algebraic expressions. 
In contrast, the higher flexibility of the current OPP-based
codes~\cite{vanHameren:2009dr,Hirschi:2011pa,Cullen:2011ac}
comes at the price of a lower CPU efficiency.
This motivates us to introduce a new one-loop algorithm that naturally
adapts to tensor-integral and OPP reduction and maximises speed and flexibility
in a way that does not depend on the employed reduction.
Inspired by the observation that colour-ordered multi-gluon amplitudes
can be efficiently computed by combining tensor integrals with
a one-loop Dyson-Schwinger recursion~\cite{vanHameren:2009vq},
we formulate a numerical algorithm that generates
one-loop amplitudes via {\em recursive construction of Feynman diagrams}.
As outlined in the following, the method is fully general, 
and first non-trivial applications demonstrate its high efficiency,
when combined with both tensor-integral or OPP reduction.

Leading-order transition amplitudes $\calM$ and virtual 
NLO corrections $\delta\calM$ are handled as sums
of tree and one-loop Feynman diagrams,
\begin{eqnarray}
\label{eq:amplitudes}
\calM 
= 
\sum_{d} \calM^{(d)}
,\qquad
\delta\calM 
= 
\sum_{d} \delta\calM^{(d)}
.
\end{eqnarray}
The corresponding scattering probability densities $\calW$
and virtual one-loop corrections $\delta\calW$ 
are 
\begin{eqnarray}
\label{eq:W}
\calW
=
\sum_{\hel,\col}|\calM|^2
,\qquad
\delta\calW 
=
\sum_{\hel,\col}2\,\re \left(\calM^*\delta\calM \right)
.
\end{eqnarray}
The sums run over colour and helicity states of each external
particle. Colour sums are performed at zero cost by
exploiting the {\em factorisation} of individual diagrams into colour factors $\calC^{(d)}$
and colour-stripped amplitudes 
\begin{eqnarray}
\calM^{(d)}
=
\calC^{(d)}\calA^{(d)}
,\qquad
\delta\calM^{(d)}
=
\calC^{(d)}\delta\calA^{(d)}
.
\end{eqnarray}
Algebraic reduction of the colour factors to a standard basis 
$\{\calC_i\}$ permits to encode all colour sums in the 
matrix
$
\calK_{ij}=
\sum_{\col}\calC_i^* \calC_j
$ 
,
which is computed only once per process (see~\cite{Bredenstein:2010rs} for details).

Colour-stripped tree diagrams $\calA^{(d)}$ are computed by a numerical algorithm that
recursively merges sub-trees. 
We call a sub-tree a subdiagram obtained by cutting a tree.
Sub-tree amplitudes are complex n-tuples $w^\beta(i)$, 
where $\beta$ is the spinor or Lorentz
index of the cut line. The label $i$ represents the 
topology, momentum and particle content of the sub-tree.
Sub-trees are recursively merged by connecting their cut lines to 
vertices and propagators:
\vspace*{-4mm}
{
\unitlength 0.75pt
\SetScale{0.75}
\begin{eqnarray}
\label{eq:LOrec}
w^\beta(i)&=&
\vcenter{\hbox{\treereclhs}}
=
\vcenter{\hbox{\treerecrhs}}
.
\end{eqnarray}
\vspace{-4mm}}

\noindent
The sub-trees $i$, $j$, and $k$ involve off-shell momenta,
but in contrast to off-shell currents
they represent individual topologies.
Cut lines are marked by dots,
and external lines are not depicted.
For brevity, quartic vertices are not shown explicitly,
but their inclusion is straightforward.
In terms of n-tuples,
the recursion step 
reads
\begin{eqnarray}
\label{eq:treerec}
w^\beta(i)&=&
\frac{\X_{\gamma\delta}^{\beta}(i,j,k)\; w^\gamma(j)\; w^\delta(k)
}{p_i^2-m_i^2+\ri\varepsilon},
\end{eqnarray}
where $X_{\gamma\delta}^{\beta}/(p_i^2-m_i^2+\ri\varepsilon)$
describes a vertex connecting $i$, $j$, $k$, and
a propagator attached to $i$.
The recursion starts with 
the external lines of a tree,
i.\,e.~the wave functions of the scattering particles, 
and terminates when the generated sub-trees
permit to build all tree diagrams.
The algorithm is based on numerical routines
that implement all wave functions, propagators and vertices.
These building blocks depend only on the theoretical model 
and are easily obtained from its Feynman rules.
This approach is similar to the tree algorithm 
implemented in {{ \sc MadGraph}}~\cite{Alwall:2007st}.
Its strength lies in 
the efficiency of colour sums
and the systematic {\em  recycling of sub-trees}
appearing in different diagrams.

Let us now consider one-loop amplitudes. A 
colour-stripped $n$-point loop diagram 
is an ordered set
of $n$ sub-trees,
$\calI_n=\{i_1,\dots,i_n\}$, connected by loop propagators:
{
\unitlength 0.75pt
\SetScale{0.75}
\begin{eqnarray}
\label{eq:npointloop}
\delta \calA^{(d)} 
=
\int
\frac{\rd^Dq\; \calN(\calI_n;q)}{D_0D_1\dots D_{n-1}}
&=&\vcenter{\hbox{\npointloop}}
.
\end{eqnarray}
\vspace{-1mm}}

\noindent
The ordering $\{i_1,\dots,i_n\}$ of the external
sub-trees in \refeq{eq:npointloop} describes the topology of this particular one-loop Feynman
diagram, independently of the coloured or colourless nature of the external
particles.  Since we do not apply any ordering selection, like e.\,g.~colour
ordering, the full set of one-loop diagrams includes all orderings
(topologies) that are allowed by the Feynman rules.
The denominators $D_i=(q+p_i)^2-m_i^2+\ri\varepsilon$ 
depend on the loop momentum $q$, external momenta $p_i$, and internal 
masses $m_i$. All other contributions from loop propagators, vertices, and external sub-trees
are summarised in the numerator,
which is a polynomial of degree $R\le n$ in the loop momentum,
\begin{eqnarray}
\label{eq:Npoly}
\calN(\calI_n;q)&=& \sum_{r=0}^R \calN_{\mu_1\dots \mu_r}(\calI_n)\; q^{\mu_1}\dots q^{\mu_r}.
\end{eqnarray}
Momentum-shift ambiguities are eliminated 
by setting $p_0=0$. This singles out 
the $D_0$ propagator, and  the loop momentum $q$ flowing through 
this propagator 
is marked by an arrow in~\refeq{eq:npointloop}.
In traditional one-loop calculations, 
the coefficients $\calN_{\mu_1\dots\mu_r}$  are 
explicitly constructed from the Feynman rules,
and the  amplitude~\refeq{eq:npointloop} is expressed as a linear combination
\begin{eqnarray}
\label{eq:ampTIcont}
\delta \calA^{(d)}&=& \sum_{r=0}^R \calN_{\mu_1\dots \mu_r}(\calI_n)\; T_{n,r}^{\mu_1\dots \mu_r}
\end{eqnarray}
of tensor integrals
\begin{eqnarray}
\label{eq:TI}
%
T_{n,r}^{\mu_1\dots \mu_r}=
\int
\frac{\rd^Dq\; q^{\mu_1}\dots q^{\mu_r}}{D_0D_1\dots D_{n-1}}.
\end{eqnarray}
These latter are subsequently reduced to $m$-point scalar integrals $T_{m,0}$
with $m=1,2,3,4$, which originate from~\refeq{eq:TI} by cancelling the
numerator and at least $n-4$ denominators $D_i$.  
Alternatively, the OPP method~\cite{Ossola:2006us} permits to by-pass
tensor integrals through a 
direct connection 
between the numerator $\calN(\calI_n;q)$ 
and the scalar-integral representation of the amplitude.
To this end, the  numerator is
expressed as a polynomial in the denominators $D_i$. The scalar-integral  coefficients are
determined by evaluating $\calN(\calI_n;q)$ at loop momenta $q$ that satisfy
multiple-cut conditions of type $D_{i}=D_j=\dots=0$. 

In this framework, the numerator can be computed with tree-level techniques.
Let us consider the {\em cut loop} 
that results from~\refeq{eq:npointloop}
by cutting the $D_0$ propagator and removing
denominators,
\vspace{2mm}
{
\unitlength 0.75pt
\SetScale{0.75}
\begin{eqnarray}
\label{eq:looprec}
\calN_\alpha^\beta(\calI_n;q)=
\vcenter{\hbox{\loopreclhs}}
&=&
\vcenter{\hbox{\looprecrhs}}
.
\quad
\end{eqnarray}
\vspace{-2mm}}

\noindent
The indices $\alpha$ and $\beta$ are associated with the arrows that 
mark the ends of the cut line, and the 
trace of the cut loop corresponds to the numerator 
$\calN(\calI_n;q)$.
As depicted in~\refeq{eq:looprec}, $n$-point cut loops 
can be  constructed by recursively merging
lower-point cut loops and sub-trees.
More explicitly,
\begin{eqnarray}
\label{eq:looprecb}
\calN_\alpha^\beta(\calI_n;q)
=
\X_{\gamma\delta}^{\beta}(\calI_n,i_{n},\calI_{n-1})
\;\calN^\gamma_\alpha(\calI_{n-1};q)
\; w^\delta(i_n),
\;
\end{eqnarray}
where $\X^\beta_{\gamma\delta}$  and $w^\delta$ are the
same vertices and sub-trees that enter the tree algorithm.
It is thus 
possible, within the OPP framework, to reduce the calculation of scalar-integral coefficients 
to a tree-level problem.
Highly automatic tree generators can be upgraded to 
loop generators~\cite{vanHameren:2009dr,Hirschi:2011pa},
thereby reducing the human power needed for NLO calculations by orders of magnitude.
However, when applied to non-trivial processes,
this approach can require massive computing resources.
The reason is that OPP reduction requires
repeated evaluations of 
$\calN(\calI_n;q)$ for a large number of $q$-momenta.

This is related to the nature of loop calculations,
which requires the 
knowledge of the numerators
as {\it functions} of the loop momentum $q$.
It is thus natural to introduce a new kind of 
loop-generator algorithm, where 
the building blocks of the recursion~\refeq{eq:looprecb} are handled as functions of
$q$. To this end, we express the cut loop~\refeq{eq:looprec} as  a
polynomial
\begin{eqnarray}
\label{eq:openloop}
\calN^\beta_\alpha(\calI_n;q)= \sum_{r=0}^R \calN^\beta_{\mu_1\dots \mu_r;\alpha}(\calI_n)\; q^{\mu_1}\dots q^{\mu_r}.
\end{eqnarray}
To emphasise the  loop-momentum dependence
encoded in the set of 
coefficients $\calN^\beta_{\mu_1\dots\mu_r;\alpha}(\calI_n)$,
we call this representation an {\em open loop}.
In renormalisable 
gauge theories, 
splitting the $X$ tensor in~\refeq{eq:looprecb} into 
a constant and a linear part,
\begin{eqnarray}
\label{eq:coeffvert}
&&\X_{\gamma\delta}^{\beta}
=
\Y_{\gamma\delta}^{\beta}
+
q^\nu\; \Z_{\nu;\gamma\delta}^{\beta}
,
\end{eqnarray}
we obtain recursion relations for $n$-point open loops
in terms of lower-point open loops and sub-trees:
\begin{eqnarray}
\label{eq:coefflooprec}
&&\calN_{\mu_1\dots\mu_r;\alpha}^\beta(\calI_n)=
\left[
\Y_{\gamma\delta}^{\beta}
\; 
\calN^\gamma_{\mu_1\dots\mu_r;\alpha}(\calI_{n-1})
\right.
\nonumber\\
&&{}+\left.
\Z_{\mu_1;\gamma\delta}^{\beta}
\; 
\calN^\gamma_{\mu_2\dots\mu_r;\alpha}(\calI_{n-1})
\right]\; w^\delta(i_n).
\end{eqnarray}
The number of coefficients grows
with the polynomial degree, which 
corresponds to the tensorial rank $r$.
However, symmetrising open-loop tensorial 
indices $\mu_1\dots\mu_r$
keeps the number of components well under control~\cite{vanHameren:2009vq}.
Once the coefficients are known,
multiple evaluations of the polynomial~\refeq{eq:Npoly} 
can be performed at a negligible CPU cost~\cite{Heinrich:2010ax}. This
strongly boosts OPP reduction.
Moreover, the same coefficients can be used 
for a tensor-integral representation of the loop amplitude~\refeq{eq:ampTIcont}.
Open loops can thus be interfaced with both OPP and tensor-integral reduction 
in a natural way.

\begin{figure}[h]
\vspace*{-0.5ex}
\begin{center}
  \includegraphics[scale=0.75]{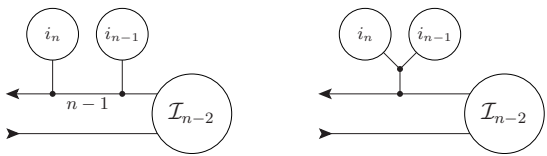}
\end{center}
\vspace{-3.5ex}
\caption{Parent (left) and child (right) open loops.
}
\label{fig:parentchild} 
\end{figure}

The efficiency of the open-loop recursion is further increased by means of
relations that arise from {\em pinching loop propagators}. Let us consider
the  parent ($n$-point) and child
($(n-1)$-point) diagrams in \reffi{fig:parentchild},
where the child results from pinching the $D_{n-1}$ propagator of the 
parent. It is evident that the parent 
can be constructed by {\em recycling} the $\calI_{n-2}$ part of the child.
But this requires that  parent and child are cut 
as in \reffi{fig:parentchild}.
To this end we order the external sub-trees 
using a function $i_k\to\cs(i_k)$ that
fulfills 
$\cs(i_k)>0$; 
$\cs(i_k)\neq\cs(i_l)$ if
$i_k$ and $i_l$ contain different external legs;
$\cs(i_k\oplus i_l)> \mathrm{max}\{\cs(i_k),\cs(i_l)\}$
where $i_k\oplus i_l$ is the merged sub-tree resulting from $i_k$ and $i_l$. 
The position and direction of the cut are determined by 
selecting contiguous sub-trees $i_1$ and $i_n$ with 
%
%
%
%
%
\begin{eqnarray}
\label{eq:cut}
\cs(i_k)> \cs(i_1)\quad \forall\quad k > 1,
\qquad 
\cs(i_n)> \cs(i_2).
\end{eqnarray}
This guarantees that parent and child diagrams are cut as in \reffi{fig:parentchild},
so that each parent can be constructed from the $\calI_{n-2}$ part of a previously
computed child.

The possibility of {\em highly efficient 
helicity sums}
is another key feature of open loops.
Unpolarised transition probabilities 
require multiple
evaluations of the polarised amplitudes~\refeq{eq:npointloop}.  The number of
helicity configurations grows exponentially with the particle multiplicity,
and the resulting CPU cost can be very large.
This can be avoided  by exploiting the 
decomposition~\refeq{eq:ampTIcont} into
helicity-dependent coefficients $\calN_{\mu_1\dots\mu_r}$
and helicity-independent tensor integrals.
The CPU expensive evaluation of tensor integrals~\refeq{eq:TI} is performed only once, and helicity sums---when
restricted to the coefficients---become very fast.
More explicitly, the contribution of~\refeq{eq:ampTIcont}
to the unpolarised transition probability is handled as a linear combination
\begin{eqnarray}
\label{eq:WTIcont}
\delta \calW^{(d)}
=
\re\left[\sum_{r=0}^R \delta\calW^{(d)}_{\mu_1\dots \mu_r}\; T_{n,r}^{\mu_1\dots \mu_r}\right],
\end{eqnarray}
with helicity- and colour-summed coefficients
\begin{eqnarray}
\label{eq:deWcoeff}
\delta\calW^{(d)}_{\mu_1\dots \mu_r}
=
2\sum_{\hel}
\left(\sum_\col\calM^*\calC^{(d)}\right)
\calN_{\mu_1\dots \mu_r}(\calI_n).
\end{eqnarray}
The unpolarised representation~\refeq{eq:WTIcont} 
can be reduced to scalar integrals with any method, including OPP.
Within the OPP framework, 
the reduction is performed by starting from the
unpolarised numerator function
$\delta\calW^{(d)}(\calI_n;q)=\sum_r \delta\calW^{(d)}_{\mu_1\dots\mu_r} q^{\mu_1}\dots q^{\mu_r}$;
in this way open loops lead to extremely fast helicity sums as compared to 
traditional tree generators. 
The OPP reduction is further improved by combining sets of 
loop diagrams with identical loop propagators but
different external sub-trees.

As a proof of concept,
we realised a fully automatic 
generator of QCD corrections to
Standard-Model processes.  Diagrams are generated with
{{\sc FeynArts}}~\cite{Hahn:2000kx};  sub-tree and open-loop topologies are
processed by a {{\sc Mathematica}} program, which concatenates them in a
recursive way, reduces colour factors, and returns {{\sc Fortran\,90}}
code.
The reduction to scalar integrals is performed in terms of tensor integrals
and, alternatively, with the OPP method.  For tensor integrals we use
{{\sc Collier}}, a private library by A.~Denner and S.~Dittmaier,
which implements the scalar integrals of Ref.~\cite{Denner:2010tr} and
reduction methods that avoid instabilities from spurious singularities~\cite{Denner:2005nn}.  
%
%
OPP reduction is performed with {{\sc CutTools}}~\cite{Ossola:2007ax} and, alternatively, 
with {{\sc Samurai}}~\cite{Mastrolia:2010nb}.
Ultraviolet and infrared divergences are dimensionally regularised.
While loop denominators are consistently treated in $D=4-2\varepsilon$
dimensions, the momenta $q^\mu$ and the coefficients $\calN_{\mu_1\dots\mu_r}$ in~\refeq{eq:Npoly}--\refeq{eq:TI} are  handled in $D=4$. 
Their $D-4$ dimensional contributions, which
yield so-called $R_2$ rational terms, are restored via
process-independent counterterms~\cite{Draggiotis:2009yb}
using the tree generator.

\begin{figure}
\hspace*{-1.5em}
{\includegraphics[width=.44\textwidth]{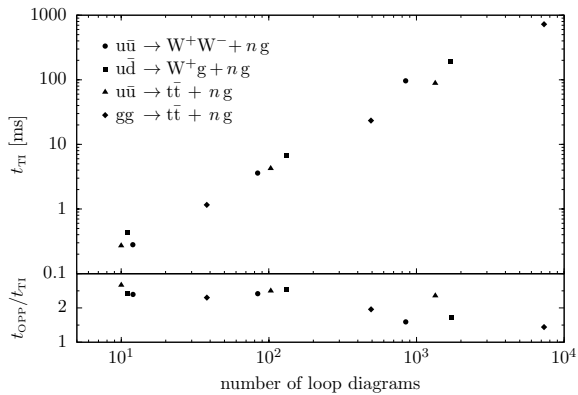}}
\caption{CPU cost of colour and helicity summed one-loop probabilities $\delta \calW$
versus number of diagrams.
Runtimes per phase space point, with tensor-integral $(t_\mathrm{TI}$) and
OPP  reduction ($t_\mathrm{OPP}$),
on a single Intel i5-750 core
with ifort 10.1.}
\label{fig:timings}
\vspace{-1.mm}
\end{figure}

To assess flexibility and performance of the method, we considered the
 $2\to 2,3,4$ reactions 
$\Pu\bar\Pu\to\PW^+\PW^-+n\Pg$, 
$\Pu\bar\Pd\to\PW^+\Pg+n\Pg$,
$\Pu\bar\Pu\to\Pt\bar\Pt+n\Pg$, and
$\Pg\Pg\to\Pt\bar\Pt+n\Pg$,
with $n=0,1,2$ gluons.  This covers
all non-trivial processes of the Les Houches
priority list~\cite{Binoth:2010ra}.  The open-loop approach leads to 
compact codes and fast code generation.
For instance---as compared to Ref.~\cite{Denner:2010jp}---the numerical 
code for \eqintext{$\Pg\Pg\to\PW^+\PW^-\Pb\bar\Pb$}
becomes two orders of magnitude smaller, 
and its generation time goes down from more than 1 week to 
4 minutes.
Also the CPU speed of open loops,
when compared against the high performance of Refs.~\cite{Bredenstein:2009aj,Denner:2010jp},
reveals a further improvement.
The CPU cost of one-loop scattering probabilities is plotted 
versus the number of diagrams in~\reffi{fig:timings}.
Sums over colours and helicities are always included.
For W bosons and top quarks,
assuming decays into massless left-handed fermions, we include a single helicity.
For the 12 considered processes, involving $\calO(10)$ to
$\calO(10^4)$ diagrams, the CPU cost scales almost linearly 
with the number of diagrams. This unexpected feature
indicates that the increase of tensorial rank 
does not represent an additional penalty at large particle 
multiplicity.
With tensor-integral reduction (upper frame),
the runtime per phase-space point
is typically below \mbox{1\,ms} for $2\to 2$ processes;
for the most involved $2\to 4$ process 
it never exceeds one second.
%
%
The ratio of timings obtained with {{\sc CutTools}}  and tensor integrals
(lower frame) shows that,
when combined with open loops, OPP reduction 
permits to achieve similarly high speed. 
While always slightly lower, the relative OPP efficiency 
seems to improve with particle multiplicity.
This holds also for {{\sc Samurai}}.

\begin{figure}
\hspace*{-2.5em}
{\includegraphics[width=.44\textwidth]{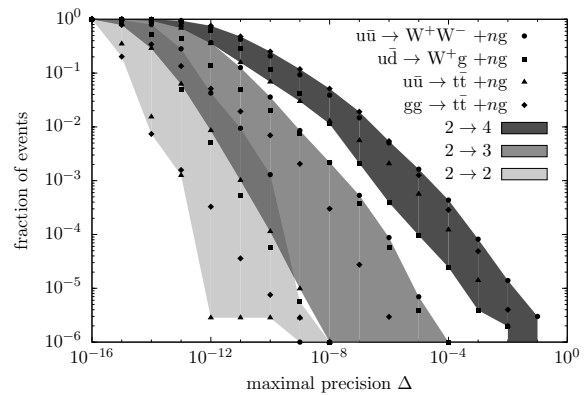}}
\caption{
%
%
Accuracy of $\delta \calW$  using tensor reduction in double precision. 
The probability of accuracy worse than 
$\Delta$, in samples of $10^6$ uniformly distributed 
phase-space points with $\sqrt{s}=1\TeV$, $p_\rT>50 \GeV$, $\Delta R_{ij}>0.5$,
is plotted versus $\Delta$.
}
\label{fig:stability}
\end{figure}

The correctness of the results is verified by comparing tensor-integral
versus OPP reductions, and checking ultraviolet and infrared cancellations.
To assess numerical instabilities, we surveyed the dimensional scaling 
of probability densities, $\delta\calW \to \xi^K \delta\calW$, with respect to $\xi$-variations  
of mass units.
Results obtained with tensor integrals
for the 12 considered processes are shown
in \reffi{fig:stability}.
In samples of $10^6$ phase space points, the average number of correct
digits for $\delta\calW$ ranges from $11$ to~$15$. For the most involved processes, 
precision lower than $10^{-5}$ and $10^{-3}$ occurs with less than
2 and 0.1 permille probability, respectively.
This demonstrates the robustness of 
the tensor-reduction approach~\cite{Denner:2005nn}
in double precision.
In contrast, with OPP reduction, a small but non-negligible fraction of 
points are not sufficiently stable in double precision. 
A detailed discussion of this aspect,
including
possible use of quadruple precision or numerical interpolation,
is deferred to a forthcoming paper.

In summary, promoting tree generators to open-loop algorithms,
we developed a fully flexible, very fast, and numerically stable 
technique to generate one-loop corrections.
Loop momenta are separated from colour and helicity
structures in a way that naturally adapts to tensor-integral and OPP
reduction, yielding excellent CPU speed with both reductions.
Open loops have the potential to address a very wide range of problems at
high-energy colliders, ranging from $2\to 2$ scattering to
multi-particle processes with up to $\calO(10^5)$ diagrams.

\begin{acknowledgments}
We are grateful to A.~Denner, S.~Dittmaier 
and L.~Hofer for providing us with 
tensor-integral reduction libraries.
We thank T.~Gehrmann for comments on the manuscript
and acknowledge support from the SNSF.

\end{acknowledgments}

\bibliography{openloops}

\end{document}